# Protonation induced high-$T_c$ phases in iron-based superconductors evidenced by NMR and magnetization measurements

Yi Cui [a], Gehui Zhang [a], Haobo Li [b], Hai Lin [c], Xiyu Zhu [c], Hai-Hu Wen [c], Guoqing Wang [d], Jinzhao Sun [d], Mingwei Ma [d], Yuan Li [d, e], Dongliang Gong [f, g], Tao Xie [f, g], Yanhong Gu [f, g], Shiliang Li [e, f, g], Huiqian Luo [f], Pu Yu [b, e*], Weiqiang Yu [a*]

[a] Department of Physics, Renmin University of China, Beijing 100872, China
[b] State Key Laboratory of Low Dimensional Quantum Physics and Department of Physics, Tsinghua University, Beijing 100084, China
[c] Center for Superconducting Physics and Materials, National Laboratory of Solid State Microstructures and Department of Physics, Collaborative Innovation Center for Advanced Microstructures, Nanjing University, Nanjing 210093, China
[d] International Center for Quantum Materials, School of Physics, Peking University, Beijing 100871, China
[e] Collaborative Innovation Center of Quantum Matter, Beijing 100871, China
[f] Beijing National Laboratory for Condensed Matter Physics, Institute of Physics, Chinese Academy of Sciences, Beijing 100190, China
[g] University of Chinese Academy of Sciences, Beijing 100049, China

*Corresponding authors: P. Yu (yupu@tsinghua.edu.cn); W. Yu (wqyu_phy@ruc.edu.cn).

**Abstract**

Chemical substitution during growth is a well-established method to manipulate electronic states of quantum materials, and leads to rich spectra of phase diagrams in cuprate and iron-based superconductors. Here we report a novel and generic strategy to achieve nonvolatile electron doping in series of (i.e. 11 and 122 structures) Fe-based superconductors by ionic liquid gating induced protonation at room temperature. Accumulation of protons in bulk compounds induces superconductivity in the parent compounds, and enhances the $T_c$ largely in some superconducting ones. Furthermore, the existence of proton in the lattice enables the first proton nuclear magnetic resonance (NMR) study to probe directly superconductivity. Using FeS as a model system, our NMR study reveals an emergent high-$T_c$ phase with no coherence peak which is hard to measure by NMR with other isotopes. This novel electric-field-induced proton evolution opens up an avenue for manipulation of competing electronic states (e.g. Mott insulators), and may provide an innovative way for a broad perspective of NMR measurements with greatly enhanced detecting resolution.

1. Introduction

For both cuprate and iron-based superconductors, carrier doping into the parent compounds is crucial to suppress the magnetic ordering and induce high-temperature superconductivity [1-5]. However, doping in bulk materials is mostly achieved through chemical substitution upon reactions at high temperatures, a method that is constrained by the chemical solubility of the dopant. Ionic liquid (or electrolyte) gating method tunes the carrier density through electrostatic gating to achieve novel superconducting states [6-13]. A novel approach has been developed recently to use lithium ion electrolyte or conductive glass ceramics as medium to achieve the electric field induced lithiation in thin flakes [14–17]. However, all these techniques are only effective for thin material with carriers penetrating at nanometer scale, and gate voltage is required during measurements. Consequently, spectroscopic techniques, such as NMR and angle resolved photoelectron spectroscopy (ARPES), are not feasible because of the low effective volume of gated phases and the large impact of the ionic liquid.

Recently, it was discovered that ion liquid gating can also lead to oxygen and hydrogen evolution into the bulk form of a thin film [18], in which oxygen or hydrogen penetrates tens of nanometers within hours. More importantly the dopants reside in the sample permanently when the electrode voltage is switched off. This motivates us to tune the carrier density of bulk superconductors with extended gating time. In particular, for quasi-2D materials with large interlayer spacing, protons may be implanted into the interstitial sites with weak impurity scattering. If hydrogen is in a nonzero valence state, an effective charge doping effect is expected.

In this letter, we report our attempts to implant protons by this room-temperature ion gating method into bulk iron-based superconductors with most common structures, including the 11 and the 122 structures. We found that protons are successfully doped into $FeSe_{0.93}S_{0.07}$ single crystals over a macroscopic length scale, whose $T_c$ is enhanced from 9 to 42.5 K. For FeS, protonation enhances $T_c$ from 4 to 18 K. These new protonated structures enables proton NMR measurements to give evidences for bulk superconductivity. Therefore, protons serve not only as a dopant for carrier doping, but also as a sensitive local probe for superconductivity, which is impossible for non-protonated compounds like FeS. To prove the generic nature of this approach, we also demonstrated that proton implantation induces superconducting for undoped non-superconducting $BaFe_2As_2$. We believe that our protonation methodology may be applied to a wide range of insulators, to tune for metal-insulator transitions, unconventional superconductivity, magnetism, etc., as well to allow for rich spectroscopy studies.

2. Techniques

The protonation to the iron-based superconductors was performed by the setup shown in Fig. 1a, following Ref. [18]. The ionic liquid is filled in a plastic container. Two platinum electrodes are separated by about 15 mm and connected to a 3 V voltage source. The single crystals were then attached to the negative electrodes by silver paint. Typical protonation peroid is about six days. We tested two types of ionic liquids, DEME-TFSI and EMIM-BF$_4$, and each producing the same enhancement of $T_c$. The single crystal X-ray data were performed on an XRD machines using Cu $\alpha$ and $\beta$ lines. The magnetization was measured by a VSM-SQUID (Quantum Design) down to 2 K. The ac susceptibility is deduced from the shift of the resonance frequency of an untuned NMR circuit by $f = f_0 (1+\chi_{ac})^{-1/2}$, where $\chi_{ac}$ is the ac susceptibility of the sample. The proton NMR spectra were collected by the $\pi/2$-$\tau$-$\pi$ spin-echo pulse sequences, with $\pi/2$ pulse length of 0.3 μs and the inter-pulse delay $\tau$ of 8 μs. The proton spin-lattice relaxation time $^1T_1$ is measured by the standard inversion-recovery method.

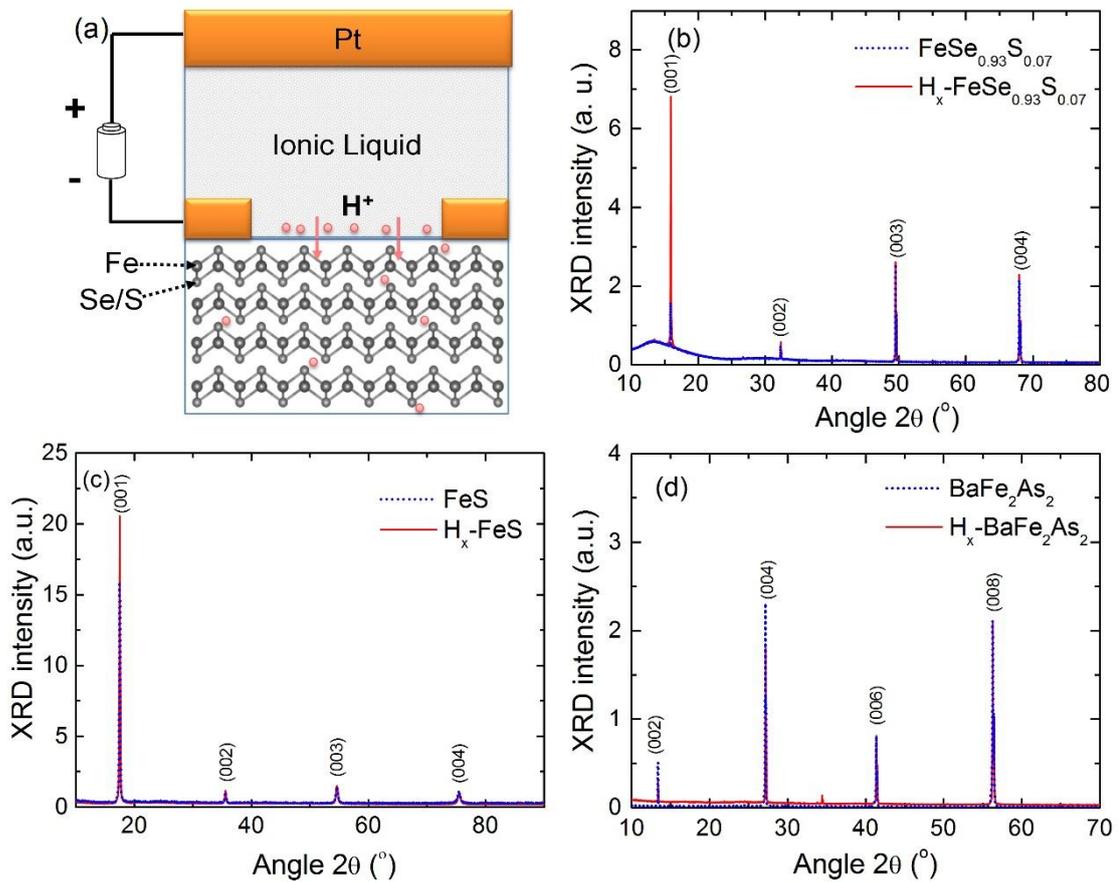

Fig. 1 (Color online) Sample protonation and structural characterization. (a) The configuration for protonation. Two parallel Pt electrodes are placed in the ionic liquid with a distance of 15 mm, applied with a 3V voltage difference. The sample is attached close to the negative electrode. (b) The single crystal X-ray diffraction (XRD) data of a FeSe$_{0.93}$S$_{0.07}$ single crystal (Sample S1, size ~ 3.24 mm×1.54 mm×0.28 mm) before and after six-day protonation. (c) The XRD data of a FeS single crystal (Sample S2, 2 mm×2 mm×0.1 mm), before and after protonation. (d) The XRD data of a BaFe$_2$As$_2$ single crystal (Sample B1, 2 mm×4.4 mm×0.2

mm) before and after protonation.

## 3. Data analysis and discussion

Since these compounds have layered structure, protons are likely inserted into interstitial sites. The XRD patterns for three single crystals, FeSe$_{0.93}$S$_{0.07}$ (Sample S1), FeS (Sample S2) and BaFe$_2$As$_2$ (Sample B1), all after six days of protonation, are shown in Fig. 1b-d. The positions of all XRD peaks do not show an appreciable change after protonation, which suggests that the bulk crystal remains undamaged. Due to the small ratio of the protonated region (typically with micrometers thickness as discussed blow), the change of the lattice constant cannot be resolved from the current XRD method and would be an interesting problem for future studies.

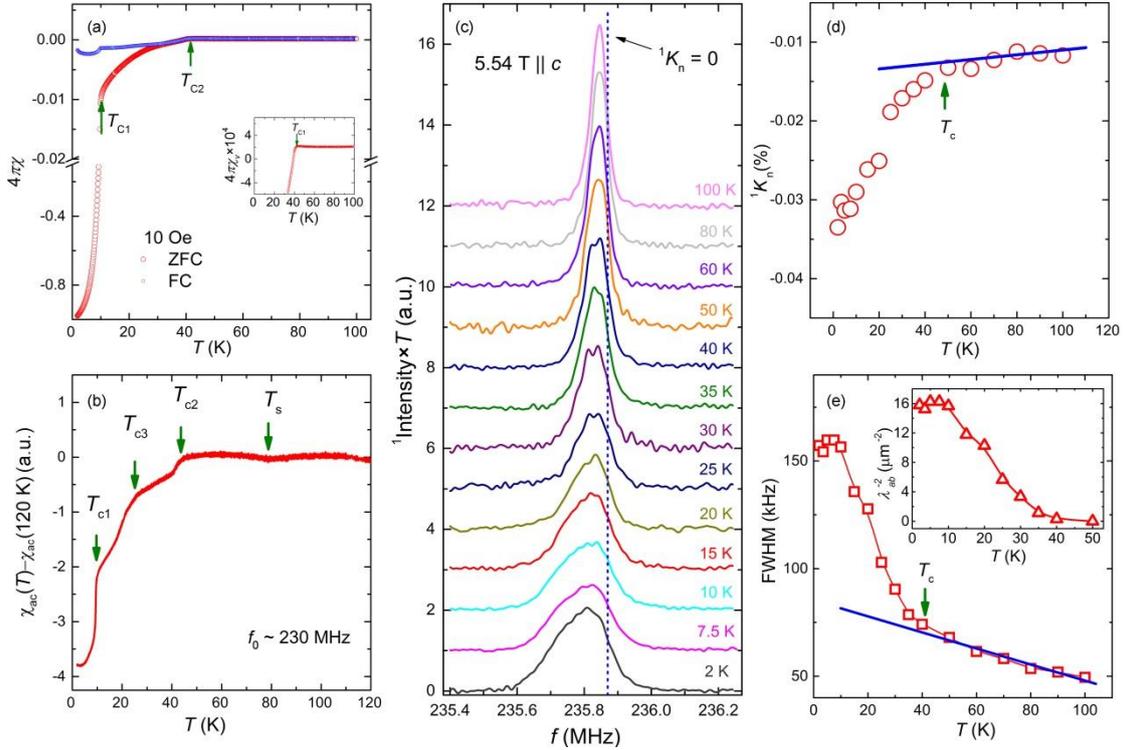

Fig. 2 (Color online) Magnetization and NMR measurements on H$_x$-FeSe$_{0.93}$S$_{0.07}$ (Sample S1). (a) The magnetic susceptibility of the sample $\chi_v$ measured under zero-field-cooled (ZFC) and field-cooled (FC) conditions with a field of 10 Oe. Two superconducting transitions are shown by the sharp decrease of $\chi_v$ with the transition temperature marked as $T_{c1}$ ($\approx$ 9 K) and $T_{c2}$ ($\approx$ 42.5 K). (b) The ac susceptibility measured by the NMR coil with Sample S1 inside, which shows the structure transition and three superconducting transitions by kinked features. (c) The $^1$H NMR spectra at different temperatures under a fixed field of 5.54 T applied along the c-axis. Data at different temperatures are offset for clarity. (d) The Knight shift $^1K_n$ as a function of temperature, with $T_c$ marking the onset temperature of superconductivity. (e) The full-width-at-half-maximum (FWHM) of the NMR spectra. Inset: the $\lambda_{ab}^{-2}$ as a function of temperature, where $\lambda_{ab}$ is the in-plane penetration depth.

We first show protonation effect in the 11 structure iron-based superconductor FeSe$_{0.93}$S$_{0.07}$ [19, 20], as an example of enhanced $T_c$ and sensitive NMR studies. The superconducting transition of the protonated sample is determined by both the dc and the ac susceptibility measurements shown in Fig. 2a and b. In Fig. 2a, the dc susceptibility $\chi_v$ under ZFC shows a primary drop at $T_{c1} \sim 9$ K, which reproduces the superconducting transition of the non-protonated sample. However, a second sudden drop occurs at a higher temperature of 42.5 K, as shown in the inset, which clearly demonstrates a superconducting transition due to superconducting diamagnetic effect. Just above 9 K, the $4\pi\chi_v$ drops to 1% of full diamagnetization ($4\pi\chi_v = -1$). Assuming that protons are primarily doped on the top layers of the sample, a thickness of 2.8 μm is estimated from the $\chi_v$ of this high-$T_c$ phase. By comparison, the ac susceptibility $\chi_{ac}$, measured by the NMR coil at high frequencies (see Sect. 2), exhibits four kinked features as shown in Fig. 2b. The onset temperatures for these features are labeled as $T_s$, $T_{c1}$, $T_{c2}$ and $T_{c3}$ (~ 25.5 K), respectively. The first three temperatures are consistent with the structural transition of undoped compounds [20] and two superconducting transitions revealed by dc $\chi_v$ in Fig. 2a. The additional rapid drop of $\chi_{ac}$ at $T_{c3}$ suggests that hydrogen doping induces multiple superconducting transitions. In fact, the three superconducting transitions at $T_{c1}$, $T_{c3}$ and $T_{c2}$ are consistent with the gate-controlled thin flakes of FeSe with the gate voltage of 0, 4.25 and 5.5 V respectively, where a doping level of 0.12 e/Fe is suggested for the $T_c \sim 42.5$ K phase [15]. This similarity of high-$T_c$ phases of both techniques supports that our protonation induces electron doping to the system, and protons remain in the 1+ valence state.

Fig. 2c-e presents the proton NMR study measured under a magnetic field of 5.54 T applied along the *c*-axis of the crystal. The proton spectra at different temperatures are shown in Fig. 2c. The spectrum has a FWHM about 50 kHz at 100 K, and grows broader upon cooling. The NMR Knight shifts $^1K_n$, calculated from the frequency of the spectral peaks, are shown in Fig. 2d, where $^1\gamma = 42.5759$ MHz/T is the gyromagnetic ratio of protons. A rapid drop of the Knight shift is seen below 40 K, coinciding with the $T_c$ measured by the magnetization data. The drop of $^1K_n$ below $T_c$ is consistent with the onset of singlet paring of superconductivity. The superconducting transition is further evidenced by the FWHM shown in Fig. 2e, which exhibits a rapid increase below 40 K. This NMR line broadening below $T_c$ is caused by the distribution of internal field in the superconducting states due to formation of the vortex state [20]. The in-plane penetration depth $\lambda_{ab}$ is then calculated by the increased NMR linewidth [21] and shown in the inset of Fig. 2e. $\lambda_{ab}$ tends to saturate below 10 K and reads about 0.25 μm at $T = 2$ K. This value is comparable to Ba(Fe$_{1-x}$Co$_x$)$_2$As$_2$ [22, 23], and in the same order as undoped FeSe ($\lambda_{ab}$~0.406 μm) [24]. Our NMR study on non-protonated FeSe$_{0.93}$S$_{0.07}$ shows an almost identical $\lambda_{ab}$ as FeSe (data not shown). With the superfluid density $n_s \propto \lambda^{-2}_{ab}$ [25], a slight increase of $n_s$ is suggested after protonation. We should emphasize that $^{57}$Fe and $^{77}$Se NMR are not feasible to study the high-$T_c$ phase in this compound, because of their much lower natural abundances and lower gyromagnetic ratios compared to proton.

For a more prominent case, we demonstrate next that protonation in FeS enhances the $T_c$ from 5 to 18 K. Fig. 3a shows the *in situ* ac susceptibility of Sample

S2 measured in the NMR coil. Two superconducting transitions are clearly resolved from the diamagnetic effect, seen from the sharp decreasing of $\chi_{ac}$ upon cooling, which are marked by two down arrows at $T_{c1}$ and $T_{c2}$ respectively. The $T_{c1}$ (~ 5 K) corresponds to the transition temperature of non-protonated crystal. $T_{c2}$ (~ 18 K) is an emergent superconducting transition at a higher temperature, which will be confirmed by our NMR studies below. We note that we did not see the $T_{c2}$ in the $\chi_{ac}$ of non-protonated samples, which indicates that the second superconducting transition is induced by protonation. From 37 K (marked as $T^*$ in the figure) down to 20 K, a weak decrease of $\chi_{ac}$ is also shown. At present, we are not able to confirm if $T^*$ is related to a superconducting or other transition.

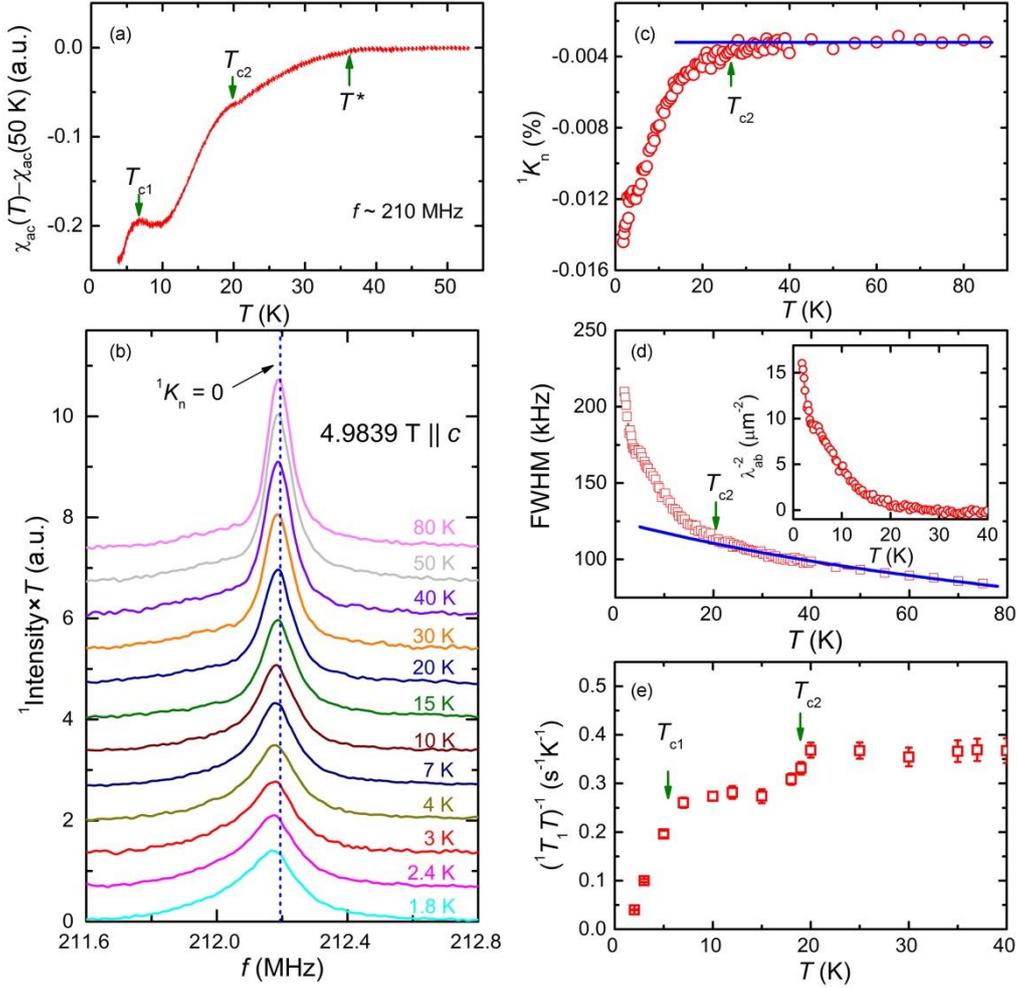

Fig. 3 (Color online) NMR study on $H_x$-FeS (Sample S2). (a) The ac susceptibility $\chi_{ac}(T)$ as a function of temperature, where $T_{c1}$ and $T_{c2}$ mark the onset temperatures of superconductivity by the sharp drop of $\chi_{ac}$. The $T^*$ at 37 K suggests an additional phase-transition like behavior by a weak decrease of $\chi_{ac}$ upon cooling. (b) The NMR spectra measured under a constant field of 4.9839 T applied along the $c$-axis. (c) The Knight shift $^1K_n$ as a function of temperature. A sudden drop of $^1K_n$ at $T_{c2}$ ~ 18 K indicates the onset of a superconducting transition. (d) The FWHM of the spectra as a function of temperature. The high temperature data are fit with a Curie-Weiss function with a deviation to data below 18 K, also suggests a superconducting transition at $T_{c2}$. Inset: the $\lambda_{ab}^{-2}$ as a function of temperature derived from FWHM, where the

Curie-Weiss contribution is subtracted. (e) The spin-lattice relaxation rate $1/^1T_1T$ as a function of temperature, where $T_{c1}$ and $T_{c2}$ marks the double superconducting transitions by the drop of $1/^1T_1T$.

Proton NMR was also succeeded in FeS, overcoming the shortage of sensitive NMR isotopes for related superconductivity studies. Fig. 3b show the $^1$H NMR spectra measured with a 4.9839 T field applied along the *c*-axis of the sample. With decreasing temperature, a broadening of the NMR spectra and a shift of the peaked position to lower frequencies are seen, in particularly below 20 K. The Knight shift (determined by the peak position of the spectra) and the FWHM are deduced from the spectra, and shown in Fig. 3c and 3d respectively. The Knight shift $^1K_n(T)$ exhibits a prominent decrease when cooled below 20 K, again consistent with superconducting transition. The FWHM can be fitted with a Curie-Weiss function FWHM $\sim a/(T+150$ K) for temperatures above 20 K, which suggests a weak paramagnetic contribution frequently seen in magnetic materials. However, the FWHM data deviates from this fitting below 20 K, and demonstrates a large enhancement upon cooling, which is again consistent with the vortex state. The penetration depth in the *ab* plane is then estimated by subtracting the Curie-Weiss contribution from the FWHM, and shown as a function of temperature in the inset of Fig. 3d. Here a penetration depth of 0.25 μm is also estimated, which is close to that of Sample S1 (the protonated FeSe$_{0.93}$S$_{0.07}$ sample). Therefore, a similar superfluid density is also suggested for this high-$T_c$ phase. We also performed the spin-lattice relaxation measurements, with data shown in Fig. 3e. Below 20 K, two superconducting transitions are resolved at 18 and 5 K respectively by the drop of $1/T_1T$ due to gap opening of superconductivity. However, the absence of the Hebel-Slichter coherence peak at $T_{c1}$ and $T_{c2}$ makes differences from conventional *s*-wave superconductors [26, 27]. Above $T_c$, the $1/T_1T$ is nearly temperature independent, similar to that of the intercalated iron selenide K$_x$Fe$_2$Se$_2$ [28], but in sharp contrast to Curie-Weiss behavior in FeSe [29]. This constant-like behavior suggests the absence of low-energy spin fluctuations, which may be understood by the significant electron doping from intercalated K$^+$ or H$^+$. The low-energy spin fluctuations, usually arising from inter-pocket scattering or magnetic correlations, is dramatically reduced, because the hole pocket is strongly suppressed upon electron doping in FeSe-like compounds [30-32].

Finally, we report protonation effect into a typical parent, non-superconducting compounds. In Fig. 4, the magnetization data are shown for BaFe$_2$As$_2$ with the 122 structure after six days of protonation. The non-protonated BaFe$_2$As$_2$ has a magnetic transition at 140 K, and does not superconduct at the ambient pressure [33]. Fig. 4a shows the dc magnetization of protonated BaFe$_2$As$_2$ under ZFC and FC condition measured with a weak magnetic field of 10 Oe. A sharp drop of magnetization is clearly seen in the ZFC data when cooled below 10 K. Indeed, a deviation between the ZFC and FC data are already seen at a conservative estimate of $T_c \sim 20$ K, which suggests a high-$T_c$ phase with a small volume fraction of the crystal. This $T_c$ is consistence with that of Ba(Fe$_{1-x}$Co$_x$)$_2$As$_2$ under an electron doping $x=0.07$ [34].

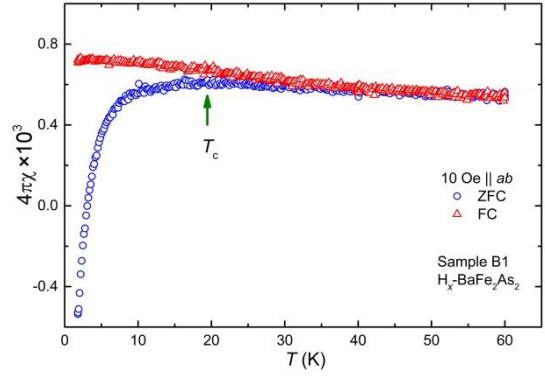

Fig. 4 (Color online) Dc magnetization $\chi_v$ on protonated $BaFe_2As_2$ (Sample B1) measured under ZFC and FC conditions. $T_c$ marks the onset temperature of superconductivity by the drop of ZFC $\chi_v$ upon cooling and the deviation between the FC and the ZFC data.

## 4. Conclusions

To our knowledge, this is the first report of $T_c$ enhancement (from 5 to 18 K) for the FeS system. Protonation in FeS leads to a high superfluid density, similar to that of $FeSe_{0.93}S_{0.07}$. The FeSe-based compounds has caused a lot of research interests lately because of their highly tunable $T_c$ [35-38]. As discussed earlier, the Hebel-Slichter coherence peak is absent in the $1/T_1$ at $T_{c1}$ and at $T_{c2}$. These observations establish an unconventional superconductivity in FeS as in $FeSe_{0.93}S_{0.07}$. This unifies the physical origin of superconductivity for all iron pnictide and iron chalcogenide superconductors.

From above measurements, we conclude that room-temperature protonation was succeeded in two typical structures of iron-based superconductors, including the 11 ($FeSe_{1-x}S_x$) and the 122 ($BaFe_2As_2$) structure. Therefore, this technique seems to be applicable for a wide range of (if not all) iron-based superconductors.The onset of high $T_c$ after protonation suggests that the impurity scattering to the 2D conducting layers by implanted protons is rather weak. By comparing with other doping methods [15], a doping level of 0.07–0.12e/Fe in $FeSe_{0.93}S_{0.07}$ is suggested after six days of protonation, which proves a large electron doping tunability by this simple method at the room temperature. We believe that this can be readily applied to other high-temperature superconductors, or more generally to quasi-2D materials, with convenient control of charge doping, to search for metal-insulator transitions and unconventional superconductivity.

On another general ground, the implanted protons allow NMR studies on materials down to micrometers due to the high sensitivity of proton NMR. In particularly, for undoped samples lacking of sensitive NMR isotopes (e.g. FeS), protonation provides an effective pathway to carry out NMR measurements. Recently, proton substitution for oxygen was succeeded in the iron-based superconductors, although the method is limited in the 1111 structure compounds and requires high-pressure synthesis [39]. We note that our technique is in an analogy to the muon spin relaxation (μSR) measurements, where $\mu^+$ are implanted into samples while in a dramatically reduced dose compared to the giant implantation level of protons in this method. It is worth noting that although our sample is only partly protonated, the proton NMR probes the doped regimes due to the existence of protons, which suggests that proton NMR holds its great advantageous as a local probe.

Finally, we want to remark that our approach makes a step toward phase control of materials, in which the ionic liquid gating induced protonation seems to be effective within a short length scale of μm after gating for a few days. Thus, for crystals with thickness over 10 μm, longer thermal equilibrium time or a higher protonation temperature may be required to obtain homogenously protonated crystal. However, for large crystals, precise control of doping is challenging, because continuous proton diffusion is unavoidable. We think that this technique may be more efficient for phase control in thin films (~100 nm), where the proton concentration can be controlled nicely by either gating voltage or gating time, as initially demonstrated in magnetic $SrCoO_{2.5}$ [18].

**Author contributions**

WY conceived the project. FeS crystals were grown with the hydrothermal method by HL, XZ and HHW. The undoped $FeSe_{0.93}S_{0.07}$ crystals were grown with the chemical vapor transport method by GQW, JZS, and MWM. The $BaFe_2As_2$ crystal were grown with the flux method by YHG, JLG, TX and HL. YC processed the protonation on all crystals with the help from PY, and performed magnetization measurements. YC, GHZ and HBL performed structural characterization. GHZ performed NMR measurements. PY and WY wrote the manuscript with inputs from all coauthors.

**Conflict of interest**

The authors declare that they have no conflict of interest.

**Acknowledgments**

The authors acknowledge discussions with Professor Mark-Henri Julien and Professor Wei Ji. This work was supported by the Ministry of Science and Technology of China (2015CB921700, 2016YFA0300504, 2016YFA0301004, 2016YFA0300401 and 2017YFA0302903) and the National Natural Science Foundation of China (11374364, 11522429, 11374011 and 11534005).